\documentclass[twocolumn,10pt,aps,prd,floatfix,nofootinbib]{revtex4}
\usepackage{latexsym}
\usepackage{psfrag}
\usepackage[dvips]{epsfig}
\usepackage{amssymb}
\usepackage{graphicx}
\usepackage{graphics}

\usepackage[dvips]{epsfig}
\usepackage{amsfonts}
\usepackage{amsmath}
\usepackage{}
\usepackage{color}
\usepackage{latexsym}
\usepackage{psfrag}
\usepackage[dvips]{epsfig}
\usepackage{amssymb}
\usepackage{graphicx}
\usepackage{graphics}

\newcommand{\ket}[1] {\mbox{$ \vert #1 \rangle $}}

\newcommand{\bra}[1] {\mbox{$ \langle #1 \vert $}}

\newcounter{subequation}[equation] \makeatletter
\expandafter\let\expandafter\reset@font\csname
reset@font\endcsname
\newenvironment{subeqnarray}
  {\arraycolsep1pt
    \def\@eqnnum\stepcounter##1{\stepcounter{subequation}{\reset@font\rm
      (\theequation\alph{subequation})}}\eqnarray}
  {\endeqnarray\stepcounter{equation}}
\makeatother

\newcommand{\ba}{\begin{eqnarray}}
\newcommand{\ea}{\end{eqnarray}}
\newcommand{\sba}{\begin{subeqnarray}}
\newcommand{\sea}{\end{subeqnarray}}

\begin{document}

\title{Problems with models of a fundamental length}
 \author{David Campo}
 \email[]{dcampo@astro.physik.uni-goettingen.de}
 \affiliation{Georg-August-Universit\"{a}t,
Institut f\"{u}r Astrophysik,
Friedrich-Hund-Platz 1, 37077 G\"{o}ttingen, Germany}
 \begin{abstract}
I critically examine various {\it ad hoc} models describing a fundamental minimal 
length at the level of the propagator.
They violate causality and/or unitarity. 
 \end{abstract}
\maketitle

A variety of thought experiments, as well as Loop Quantum Gravity and 
String Theories indicate that a minimal length $L$
manifests itself as one approaches the regime of Quantum Gravity
(see \citep{Garay} and references therein), possibly, 
but not necessarily, accompanied with a breakdown of Lorentz symmetry.

There are unfortunately few consistent models of such a minimal length 
with which we can make reliable predictions.
The case of non-commutative field theories (NCFT) is instructive.
Some of them (namely with only noncommuting space variables) can be obtained as 
limits of string theories on non-trivial backgrounds \cite{Seiberg}, 
and they seem to be unitary theories. 
By contrast it was found that NCFT who cannot be defined as a limit 
of a unitary string theory \cite{NCFT-Efield} are not consistent quantum theories  \cite{NCFTnonunitary}.

Because of this relative shortage of consistent frameworks, models are 
often formulated as {\it ad hoc} modifications of field theories.
Adopting this approach, one should keep in mind the important {\it caveat} 
illustrated by NCFT, to wit, 
although one may expect that models defined as the limit 
(e.g. low energy, weak coupling) 
of a certain unitary theory are consistent quantum theories, we
have no guarantee that an {\it ad hoc} modification of relativistic 
field theories defines a consistent model. 
Put differently, one cannot tinker with one axiom of QFT, for instance Lorentz invariance
or locality,
without considering the repercussions this might have on the whole theory, 
in particular a violation of unitarity or causality.

A few works putting forward this warning have already appeared
\cite{UVcompletion,SpeedofSound}.  The present notes are along the same lines.
I will focus on models using modified propagators, which as far as I know 
is an aspect of the problem that has not been treated in
previous studies. 

I first introduce three classes of modified propagator and show that they are not 
equivalent. Exemples of two of these classes have been proposed previously, with 
very different motivations. I do not discuss the relevance of these 
motivations but examine in sections \ref{sec: meas and DR} and \ref{sec:min L} 
the properties of the propagators 
with regards to causality and unitarity. 
I comment on an additional proposal in section \ref{sec:other}.
Finally note that, although the models considered here 
describe putative quantum gravity corrections at low energies,
they are all non gravitational.

\section{Three inequivalent classes of modified propagators}
\label{sec:def W}

\subsection{Three classes of deformed propagators}

There are (at least) three non-equivalent 
ways to deform the Lorentz invariant Wightman function.
Two of them are obvious from 
\ba \label{general W}
  W(t,r) &=& \int\!\!\frac{\mu(k) d^3k}{(2\pi)^3} \, 
  \frac{e^{-i\omega_k t + i{\bf k}{\bf x}}}{2\omega_k}
\nonumber \\
   &=& \frac{i}{4\pi^2} \int_0^{\infty}\!\!\!dk \, 
  \frac{k \mu(k)}{\omega_k} \,  e^{-i\omega_k t} \, \frac{\sin kr}{r} \qquad
\ea  
that is we can assume either a non-relativistic dispersion relation 
$\omega_k \neq \sqrt{{\bf k}^2}$ (I consider massless fields throughout),
or a non-relativistic measure $\mu({\bf k}) \neq 1$.
In both cases, the Wightman function is no longer Lorentz-invariant
(recall that $d^3k/k^0$ is the invariant measure for on-shell integrals).
The assumptions implicit in (\ref{general W}) are that the theory to which the 
propagator belongs contains a stable ground state in a prefered frame 
(hence appear only positive frequencies) which is 
translation invariant (hence the dependence on the 
difference of the positions) and isotropic (hence
the dependence on $r=\vert {\bf x} \vert$).
Both these models are briefly discussed in sec. \ref{sec: meas and DR}.

A third possibility, which appears perhaps less natural, consists
in replacing $r$ by a function $R(r)$. 
Since $R$ has the dimension of a length, it
depends on at least one additional length scale $L$.
We only require to recover the inverse square-law of 
long range forces, that is $R \sim r$ for $r \gg L$
(or $r \gg {\rm max}\left\{L_i \right\}$ in the case of 
several length scales).
All other things being equal, that is 
if $\omega = \vert {\bf k} \vert$ and $\mu({\bf k}) = 1$, 
this amounts to defining the modified Wightman function by
\ba \label{class 3}
  W = \frac{-1}{4\pi^2} \frac{1}{(t-i\epsilon)^2 - R^2(r)}
\ea
Note that there are two possible choices of $R$, 
namely $R = \sqrt{r^2 \pm L^2}$, 
such that $W$ depends on the Lorentz 
invariant distance $x^2 = t^2 - r^2$ only instead of $t$ and $r$ 
separately.
These are two candidate propagators to
describe a minimal length with Lorentz invariance. 
Their properties are described in sec. \ref{sec:min L}.

\subsection{Proof of the inequivalence}

I said that these three deformations are not equivalent.
This is obviously true for the first two, because the phase $e^{-i\omega_k t}$
cannot be cancelled by a redefinition of the measure $\mu(\bf k)$ which is real.
The nonequivalence between (\ref{class 3}) and the other two is also proved
from the expression of the Fourier transform w.r.t. the position
\ba
  W(t,{\bf k}) &=& \int\!\!d^3x \, e^{-i{\bf k}{\bf x}} \, W(t,{\bf x})
\nonumber \\
   &=& \frac{1}{i\pi k} \int_{-\infty}^{+\infty}\!\!\!dr\, 
   \frac{r e^{ikr}}{R^2 - (t-i\epsilon)^2}
\ea
where I defined the analytic continuation to negative values of $r$ by 
$R(-r) = R(r)$.
Since $k > 0$ we can close the integration contour in the upper half-plane.
In terms of $r_t^{(i)}$, the solutions of $R(r) = \vert t -i\epsilon \vert$
with ${\rm Im}(r_t^{(i)}) > 0$, 
\ba
  W(t,{\bf k}) = \sum_i \frac{r_t^{(i)}}{\vert t \vert R'(r_t^{(i)})} \, 
  \frac{e^{ikr_t^{(i)}}}{k}
\ea
For instance for $R=r$ we have a single pole in the upper half plane depending on 
the sign of $t$, namely $r_t = \pm t(1- i  \epsilon)$. 
Similarly for $R^2 = r^2 + L^2$, we have 
$r_t = \pm \sqrt{t^2 - L^2} (1- i  \epsilon)$ if $\vert t \vert \geq L$ 
and $r_t = \pm i \sqrt{L^2 - t^2}$ otherwise. 
Now, the prefactor $\frac{r_t}{t R'(r_t)}$ is real, so it cannot be identified to
a phase $e^{-i\omega_k t}$, and it depends explicitely on time, 
so it cannot be assimilated to a 
nontrivial measure $\mu({\bf k})$.
In addition, because of the nontrivial time dependence $r_t$, the
factor $e^{ikr_t}$ cannot come from a modified dispersion relation,
which completes the proof.

\section{Modified measures and dispersion relations}
\label{sec: meas and DR}

\subsection{Modified dispersion relation}

The cluster decomposition principle aside (it concerns only higher order 
correlation functions), relativistic QFTs rest on the axioms of 
microcausality and positivity of the energy-momentum spectrum in every inertial frame
\cite{bookWeinberg}.
These two conditions are sufficient to show that dispersion laws, stable particles,  
as well as thresholds of continuums are Lorentz invariant \cite{BrosEpstein}.
Hence breaking on-shell Lorentz invariance comes at the price of 
either non-causal propagation or a non stable spectrum.

Propagators characterized by a modified dispersion relation
$\omega_k \neq k$ can
be defined as the inverse of the second variation of an action.
To write the latter, two distinct approaches are possible.
One can either assume that the theory is Lorentz invariant, but that this symmetry is 
broken by a background solution. Or one can assume that
a preferred frame exists at a fundamental level, independently of 
any solution one might choose (see \cite{Mattingly} for a comprehensive review).

In the case of non gravitational low energy effective field theories described by a 
Lorentz invariant Lagrangian, 
the requirement of UV analyticity of the $S$-matrix 
forces the sign of certain leading irrelevant operators to be positive, which
in turn prohibits superluminal propagation on 
backgrounds breaking Lorentz symmetry \cite{UVcompletion}.

I am not aware of a similar analysis for models with a 
fundamental preferred frame, nor do I know of any attempt to 
UV complete such a model. 
Nevertheless one can argue the same constraints apply on the 
effective theory. For this I use the 
off-shell formulation of the model.
We can write the action in a covariant form with the help of 
a unit vector $n$ and the induced metric $q^{ab} = g^{ab} + n^a n^b$ 
on the surfaces orthogonal to $n$
\ba \label{action}
  S &=& - \frac{1}{2} \int\!\!d^4x \, \sqrt{q} \left\{ 
   (n^a \partial_a \phi)^2 
 \right. 
\nonumber \\
&& \qquad \left.
+ \,\, q^{ab}\partial_a \phi  \partial_b \phi 
   + \phi F(\Delta) \phi
 \right\}
\ea
$\Delta = q^{-1/2} \partial_a\left( \sqrt{q}q^{ab} \partial_b \right)$ 
is the corresponding Laplacian.
In the prefered frame defined by $n$, the dispersion relation reads
$\omega_k^2 = k^2 + F(-k^2)$.

Now, if gravity is turned on, the preffered frame must be defined 
dynamically in order to perserve the Bianchy identities \cite{DynPreffered}.
But without gravity, 
(\ref{action}) simply defines a non relativistic field theory, as considered 
routinely in condensed matter problems. There is dynamically no distinction possible
between a fundamental preferred frame as in (\ref{action}) or a preferred frame
defined by a non trivial background solution of a relativistically invariant theory.
Hence $n$ is a passive field and the demonstrations 
of \cite{UVcompletion} in principle also apply.
But let us keep in mind that the lack of an explicit UV completion of such a model 
is a loophole of this argumentation.

\subsection{Modified measure}

One can note that the Wightman function verifies the equation 
\ba
  \left( \partial_t^2 - \Delta \right) W(t,{\bf x}) = 0 \, .
\ea
Of course, it does not transform like a scale under boosts. 
Rather, if we boost for instance along the $x$-axis with a Lorentz factor 
$\gamma$, we get  
\ba
  W(t',{\bf x}') = \int\!\!\frac{d^3k}{2\omega_k} \, 
  \mu\left[ \gamma (k - \beta k_x) \right]
 \, \frac{e^{-i\omega_k t + i {\bf k} {\bf x}}}{(2\pi)^3 }
\ea
with $t' = \gamma(t-\beta x)$, $x' = \gamma(x-\beta t)$, $y'=y$, and $z'=z$.
The Fourier transform of the Wightman function exists, is positive, 
integrable, and vanishes outside the mass shell 
\ba \label{strange meas}
  \rho(\omega,{\bf p}) = \frac{\theta(\omega)}{(2\pi)^3} \, \mu(p^0) \delta(p^2)
\ea
Yet it is not clear (to the author) whether the model admits a consistent 
particle interpretation: if one interpretes (\ref{strange meas}) by analogy with
relativistc QFTs, it means that the one-particle spectrum is 
$\vert \bra{0} \varphi(0) \ket{{\bf p}} \vert^2 = \mu(\vert {\bf p} \vert) \, \delta(p^2)$, 
see equation (\ref{expansion rho}), which can hardly be interpreted as the spectrum 
of the energy-momentum operator. Moreover the action of $\varphi$ is nonlocal.
The model can alternately be defined by the auxillary field
\ba
  \Phi(x) \equiv \int\!\!d^4y \, \sqrt{\mu(x-y)} \, \varphi(y)
\ea
with 
\ba
  S_{\Phi} = \frac{1}{2}\int\!\!d^4x \, (\partial_a \Phi)^2  + S_{\rm int}
\ea
We can thus expect that the model is unitary, but it seems rather contrived.

\section{Minimal length with Lorentz invariance}
\label{sec:min L}

I consider the propagators (\ref{class 3}), specializing to the
two simplest representatives of this class, namely 
\ba
  G_F^\pm(x) = \frac{i}{4\pi^2} \, \frac{1}{-t^2 + {\bf x}^2 \pm L^2 + i\epsilon} \,.
\ea
Their interest lies in that they only depend on the proper length.
It is easely seen that they cannot be defined as the second 
variation of a (local or non local) action, but  
I recall that they can be introduced from a modified path integral.
I then show that their Fourier-Laplace transform cannot be interpreted as a 
spectral function. Next I study the properties of the commutator function:
$G_F^-$ is found to violate causality and to have an ill-defined density of modes.
By contrast $G_F^+$ free from these pathologies.
I then discuss the implications of these results with regard to physical predictions.

\subsection{Definition from a modified path intergral}

Padmanabhan found that the propagator
\ba \label{Gpadma}
  G_F^+(x) = \frac{i}{4\pi^2} \, \frac{1}{-t^2 + {\bf x}^2 + L^2 - i\epsilon}
\ea 
can be defined by analytic continuation from a Euclidian path integral 
with a modified measure \cite{Padmanabhan}, namely
\ba \label{K to Gpadma}
  G_E^+(x) = \int_0^{\infty}\!\!ds \, e^{- m^2 s - \frac{L^2}{4s}} K(s, x)
\ea
where $s$ is Schwinger's fifth time coordinate,
$K$ is the heat kernel solution of
$(-\partial_s + \Delta) K = \delta(s) \delta^{(d)}(x)$
and $\Delta$ is the $4$d Laplacian operator. 
In $4$ dimensions we have $K = e^{-x^2/4s}/(4\pi s)^{2}$, the expression 
(\ref{K to Gpadma}) indeed gives (\ref{Gpadma}).
Since $K$ can be interpreted as the probability amplitude for a 
particle with Hamilotian $\Delta$
to diffuse from $0$ to $x$ in a time $s$, one can say that
the factor $e^{- \frac{L^2}{4s}}$ penalizes
those paths which take a proper time shorter than $L^2$.

Similarly, one shows that the propagator 
\ba \label{Gparker}
  G_F^-(x) = \frac{i}{4\pi^2} \, \frac{1}{-t^2 + {\bf x}^2 - L^2 + i\epsilon}
\ea
is by contrast obtained from the path integral
\ba \label{K to Gparker}
  G_E^-(x) = \int_0^{\infty}\!\!ds \, e^{- m^2 s + \frac{L^2}{4s}} K(s, x)
\ea
that is by replacing the factor $e^{-L^2/4s}$ by $e^{+L^2/4s}$, thus favouring
the paths of proper time shorter than the minimal length.
Intuitively this is exactly the opposite of what we would like to 
achieve with the model of a minimal length, and we can expect to find that the 
phenomelogy of this propagator largely departs from standard physics. 
We will see that this is indeed the case. 
Comparison with (\ref{Gpadma}) will dramatically illustrate how the apparently 
innocuous change $-t^2 + {\bf x}^2 + L^2 \mapsto -t^2 + {\bf x}^2 - L^2$ can 
have drastic repercutions.

\subsection{No spectral function}
\label{sec:spectral}
The first important property of both (\ref{Gpadma}) and  (\ref{Gparker}) is that 
their Laplace-Fourier transform cannot be interpreted as a spectral function, 
that is as the spectrum of the operator $P^2$.

The Wightman function and spectral function are
the Fourier transform from one another (see for instance the chapter $10$ of \cite{bookWeinberg})
\ba
  W(t,r) 
   &=& \int\!\!d^4p \, e^{ipx} \frac{\theta(p^0)}{(2\pi)^3} \rho(-p^2) 
\ea
This expression is obtained by inserting a complete family of one particle states $\ket{n}$
in the expectation value $\bra{0} \varphi(x+y) \varphi(y) \ket{0}$. The 
one particle states are assumed to be energy-momentum eigenstates with 
momentum $p_n$, so that 
\ba \label{expansion rho}
  \frac{\theta(p^0)}{(2\pi)^3} \rho(-p^2) = \sum_n \delta(p^2-p_n^2) \, 
  \vert \bra{0}\varphi(0) \ket{n} \vert^2 \geq 0 \, .
\ea
For instance for a free field of mass $m$, we simply have $\rho(-p^2) = \delta(p^2+m^2)$. 
Equation (\ref{expansion rho}) shows explicitely that $\rho$ is positive. 
The Feynman propagator is related to $\rho(s)$ by
\ba \label{spectral}
  G_F(x^2) = \int_0^{\infty}\!\!dm^2 \, \rho(m^2) \, G_F^{(0)}(x^2; m^2) \, .
\ea
where $G_F^{(0)}(x^2; m^2)$ is the Feynman propagator of a free field of mass $m$
\ba \label{G_F}
  G_F^{(0)}(x^2; m^2) = \frac{i}{16\pi^2} \int_0^{\infty}\!\!\frac{ds}{s^2} 
 \, e^{-im^2s}e^{ix^2/4s} \, .
\ea
The positivity of $\rho$ implies that $G_F$ cannot decay faster that $1/p^2$ as 
$\vert p^2 \vert \to \infty$.
Moreover, the commutation relations expressed in the form 
$\frac{\partial}{\partial x^0} W(x-y)\vert_{x^0=y^0} 
= -i\delta^{(3)}\left( {\bf x}- {\bf y}\right)$
imply the sum rule
\ba \label{sum rule}
  \int_0^\infty\!\!dm^2 \, \rho(m^2) = 1 \, .
\ea

With these definitions and properties  
now recalled, we ask whether the modified propagators $G_F^\pm$
admits a spectral representation $\rho_\pm(m^2)$. The following results are shown in
the Appendix.
I begin with $G_F^+$. 
Its Laplace-Fourier transform is not positive, the sum rule (\ref{sum rule})
is not verified, and for $mL\ll 1$, i.e. the low energy regime 
where the propagator is assumed to describe corrections from quantum gravity, 
one finds
\ba
  \rho_+(m^2) = - \frac{L^2}{8} \left\{  1+ O\left((mL)^2 \right) \right\} \, .
\ea
We conclude that the Fourier-Laplace transform $\rho_+$ of 
(\ref{Gpadma}) cannot be interpreted as a spectral function of a local QFT
\footnote{One could wonder whether one should not instead interprete (\ref{Gpadma})
in terms of "quasi-local QFTs". In this framework,
the notion of strictly localizable observables is sacrificed \cite{quasilocal}, so it
seems {\it a priori} to be a consistent thing to do. 
It turns out that this is not necessary.
Let us remind that in quasi-local QFTs, 
the operators are not tempered distributions but are 
smeared by real analytic functions in $x$-space. This implies that the 
Wightman functions in momentum space can grow as fast as an exponental
$e^{L\vert p \vert}$, while if Wightman's axioms are adopted, 
the growth is instead polynomial. 
In our case, the Fourier transform of $1/(x_E^2 + L^2)$ is 
$\frac{L}{p} K_1(pL) \propto e^{-pL} $ for $pL \gg 1$. So quasi-local QFTs
need not be invoqued.}. 
That does not mean that there does not exist any unitary theory 
which would give (\ref{Gpadma}) as a low energy propagator. 
It means that this theory, if it exists, is neither a quantum field theory, nor
a string theory.
 In any case, we do not know what are the elemetary excitations of this theory, but we
know that they do not resemble particles as we know them.
From this point of view, one can hardly argue that (\ref{Gpadma}) 
describes "small" corrections introduced by quantum gravity.

Turning now to (\ref{K to Gparker}), we have
by contrast $\rho_- \geq  0$. But $\int\!\!dm^2\, \rho_- = \infty$, in particular 
due to an infrared divergence in the small $mL$ development of $\rho_+$
\ba
  \rho_-(m^2) \sim \frac{1}{4\pi m^2} \, .
\ea
Thus $\rho_-$ cannot be interpreted as a spectral function either.

\subsection{Causality}
\label{sec:Gpadma}

In this subsection and the next, I consider the commutator defined by
\ba \label{G_c}
  G_c^\pm(x) \equiv W^\pm(x) - W^\pm(-x) 
\ea
where the Wightman function, by definition, has the same functional 
form as (\ref{Gpadma}) and (\ref{Gparker}), but for the $i\epsilon$ prescription,
namely
\ba
  W^\pm(t,r) = \frac{1}{4\pi} \frac{1}{-(t-i\epsilon)^2 + r^2 \pm L^2}  
\ea
respectively.
Applying $1/(x\pm i\epsilon) = {\rm pv} \mp i\delta(x)$, the commutator can be written
\ba
  G_c^\pm(x) 
   &=& \frac{-i}{4\pi R^\pm}
    \left[ \delta(t - R^\pm) - \delta(t+R^\pm\right]
\nonumber \\
  R^\pm({\bf x};L) &=& \sqrt{{\bf x}^2 \pm L^2}
\ea 
and I recall that the $+$ sign is for (\ref{Gpadma}) and $-$ for (\ref{Gparker}). 

The reason for considering the commutator is threefold:
in quantum field theories, it is related to the canonical commutation relations, 
to the retarded and advanced Green functions, and to the density of modes.
We have already examined the first aspect in the previous section, where
we saw that the commutation relations 
$\frac{\partial}{\partial x^0} G_c \vert_{x^0=0} = -i\delta^{(3)}\left( {\bf x}\right) $
imply the sum rule (\ref{sum rule}), and that neither 
(\ref{Gpadma}) nor (\ref{Gparker}) verifies the latter, so they do not verify the former either.
The density of modes is the subject of the next subsection, and 
this subsection is concerned with their causal properties.

Let is begin with (\ref{Gpadma}). The support of 
$G_c^+$ is the double-sheeted hyperboloid
$ t^2 - {\bf x}^2 = L^2$ located inside the light cone. 
The commutator therefore vanishes for spacelike separated events 
$\vert t \vert < r$.
In that sense, microcausality is preserved.
Indeed, if we now define the retarded propagator by
$G_{\rm ret}(t,{\bf x}) = i \theta(t)G_c(t, {\bf x})$,
the signal from a local source 
$J(t,{\bf x}) = J_0 \delta(t) \delta({\bf x})$ is propagated according to 
\ba \label{source}
   &&\int\!\!d^4x_1 \,
   G_{\rm ret}^+(t-t_1,{\bf x}-{\bf x}_1) \, J(t_1,{\bf x}_1)
\nonumber \\
  &=& \frac{J_0}{4\pi R^+(x;L)} \theta(t) \delta\left[t-R^+(x;L)\right]
\ea
Hence the signal from the source $J$ propagates only after a lapse of time $L$
and the propagation of the "wavefront" is subluminal.

I now turn to (\ref{Gparker}). 
I recall that it differs from (\ref{Gpadma}) by the change 
$-t^2 + r^2 + L^2 \mapsto -t^2 + r^2 - L^2$.
The commutator has now its support on the spacelike hyperboloid
$x^2 -t^2 = L^2$, so microcausality is violated. 
We also see that $G_{\rm ret}^-$ does not propagate signals 
with a spatial extension smaller than $L$, no matter how long they act.

In both cases, the difference with the luminal behavior is significant
only over a region of extension $O(L)$, the ratio of the four volumes inside the 
"light-cones" scaling like $L^2/t^2$ at large times.
This violation is therefore mild and unlikely to be observable on
macroscopic scales. 
I shall not insist on these causal properties because it is not 
clear in which sense one should understand them. 
To make sense of the notion of causality for (\ref{Gpadma}) or (\ref{Gparker}), 
one would require either a field that would couple to the source $J$ in (\ref{source}),
but we know that there is no such field description of the model
(I did not write $\varphi =$ on the left hand side of (\ref{source}));
Or one would require that they are solutions of  
a partial differential equation. Its characteristics could then be interpreted as
the propagation of a pointlike source.
But there is no such partial differential equation either.

\subsection{Density of modes}

The Fourier transform of the commutator $G_c(\omega,{\bf x})$ is proportional to 
the density of modes at the frequency $\omega$. For (\ref{Gpadma}) it is
\ba \label{G_c(omega)+}
  G_c^+(\omega,{\bf x}) = e^{-\epsilon \vert \omega \vert} 
    \frac{\sin \omega R}{2\pi R}
\ea
I left the $\epsilon$ of (\ref{Gpadma}) finite 
to show that $L^2$ in (\ref{Gpadma}) is not a simple
energy cutoff, and to show that the limits $\epsilon \to 0$ and $L \to 0$ are
both well defined and commute.
The only difference with the Lorentz invariant case is the replacement of
$\vert {\bf x} \vert$ by $\sqrt{{\bf x}^2 + L^2}$. The limit $\vert {\bf x} \vert \to 0$ 
is the standard result.

The effective density of modes of (\ref{Gparker}) is very different.
We need to discuss two cases separatly.
First, for $\vert {\bf x} \vert \geq L$, the Fourier transform of the 
commutator is the same as (\ref{G_c(omega)-}), now with $\sqrt{{\bf x}^2 - L^2}$.
Second for $\vert {\bf x} \vert \leq L$ the poles are now imaginary. 
Noting $\bar R = \sqrt{L^2 - {\bf x}^2}$, the poles of 
the Wightman function are $(\pm i\bar R + \epsilon)$, hence
\ba \label{G_c(omega)-}
  G_c^-(\omega, \vert {\bf x} \vert \leq L) = \sinh(\omega \epsilon) \, 
  \frac{e^{-\vert \omega \vert \bar R}}{2\pi \bar R}
\ea
The reader can check that the limits $\epsilon \to 0$
and $L \to 0$ do not commute, that the Lorentz invariant result is not recovered
in any of the possible limiting cases, and that $G_c^-(\omega, {\bf x})$ 
has a discontinuity at $\vert {\bf x} \vert = L$.

We will now see how the expression of the density of modes modifies the 
expression of inclusive observables.

\subsection{Predictions ?}

What predictions can one make with these propagators.
We saw that neither (\ref{Gpadma}) nor (\ref{Gparker}) admit a particle interpretation. 
This alone considerably limits 
their interest since they cannot be used in calculations of $S$-matrix elements 
between asymptotic states, for there are neither asymptotic states, nor 
LSZ-reduction formula, nor cutting rules
with which to calculate scattering amplitudes and cross sections. 
This means that they can only be used to calculate vacuum-to-vacuum processes, 
exemples of which can be found in \cite{PadmaUnruh} for (\ref{Gpadma}) 
(namely the Casimir effect, 
an effective potential, an effective Lagrangian, 
and the trace anomaly).
It was found that the modified propagator produces small 
corrections suppressed by $L$ times the relevant mass scale squared.
For instance for the Casimir effect between parallel plates distant
by $a$, the relative correction is proportional to $(L/a)^2$, a very small number.  
I am not aware of similar results with the propagator (\ref{Gparker}), but 
these calculations are readily adapted, and all the 
vacuum expectation values are obtained by analytic continuation $L^2 \mapsto - L^2$.
In either case, there is no hope to test these models with such tiny corrections.

Despite this, these propagators have been considered in investigations of 
Hawking radiation and the Unruh effect \cite{Gparker}, which I now review.
This will illustrate how the properties described in the previous subsections 
transilate on physical observables, 
and in passing I will correct a glaring error found in \cite{Gparker}.

The simplest model conceivable to show that 
uniformly accelerated detectors coupled to fields in the Minkowski 
vacuum react as in a thermal bath consists of a point detector 
following a trajectory $x_{\rm det}(\tau)$ parametrized by its proper time $\tau$,  
and locally coupled to a scalar field. 
The detector possesses two internal states $\ket{\pm}$ separated by
an energy gap $E$ in the rest frame.
The interaction Hamiltonian is thus 
$H = g \left( e^{iE\tau} \ket{+} \bra{-} + h.c.\right) \varphi[x_{\rm det}(\tau)]$.
The transition rates from the ground (excited) state ($\pm$ sign) are 
given at lowest order by
\ba \label{R_pm}
  R_{\pm}(\tau) &=& 
  2g^2 Re\int_0^{\infty}\!\!d\tau_1 \, e^{\mp i E \tau_1} {\cal W}(\tau, \tau-\tau_1)
\qquad
\ea
where ${\cal W}(\tau_1, \tau_2) = \bra{0} \varphi(x(\tau_1))  \varphi(x(\tau_2)) \ket{0}$
is the Wightman function of the field evaluated at two points on the trajectory of 
the detector. 

One should keep in mind that equation (\ref{R_pm}) is derived in QFTs 
but has to be postulated "by analogy" for the propagators (\ref{Gpadma}) and 
(\ref{Gparker}). Indeed, transition rates such as (\ref{R_pm}) are obtained 
from scattering amplitudes after summing over the one-particle 
final states of the quantum field emitted or absorbed during the process.
But we saw that no such description 
is available for the propagators (\ref{Gpadma}) and (\ref{Gparker}).
Matters are even worse if we use a fully 
relativistic model where the two-level detector is replaced by 
two massive scalars fields $\Psi_M$ and 
$\Psi_m$ of masses $M > m$.
(The first quantized model is recovered in the limit $M - m \ll M$.)
In that case there are no asymptotic states to describe the detector either.

With this caveat in mind, let us calculate the transition rates (\ref{R_pm}).
We can afford ourselves to be a bit more general and consider 
two-point Wightman functions of the form 
\ba \label{W_f}
  W_f = -\frac{1}{4\pi^2} \frac{1}{f\left[ (t-i\epsilon)^2 - {\bf x}^2\right]}
\ea
which depend on the invariant length $s^2 = t^2 - {\bf x}^2$ but not on $t$ and ${\bf x}$ 
separately. I assume that $f(s^2)$ has dimension 
$\left[{\rm length}^2\right]$, and that it has an arbitrary number of simple zeros $z_k$
in the complex $z$-plane. 
Both for inertial and uniformly linearly accelerated detectors, 
the sections of $W_f$ on the detector's trajectory depend only on the difference of 
proper times. 
The transition rates (\ref{R_pm}) are therefore independent of $\tau$ and
can be furthermore calculated with the theorem of residues 
(this contrast with models characterized by a non relativistic dispersion 
relation, for which stationarity is lost \cite{UnruhLV}).
For inertial detectors we have
\ba \label{rates}
  R_{\pm} &=& \frac{-g^2}{4\pi^2}\int_{-\infty}^{+\infty}\!\!d\tau \, 
  \frac{e^{\mp iE\tau}}{f[(\tau - i\epsilon)^2]} \, .
\ea
One then readily shows that
\ba
  R_- - R_+ = g^2 \int_{-\infty}^{\infty}\!\!d\tau \, e^{iE\tau} \, G_c(\tau)
\ea
which, for inertial detectors, is proportional to the 
time-Fourier transforms (\ref{G_c(omega)+}) 
and (\ref{G_c(omega)-}) at ${\bf x}=0$ of the commutators. 
We can therefore expect that the transition rates calculated with 
(\ref{R padma}) are deformations of the standard ones, while the pathologies of 
(\ref{G_c(omega)-}) produce senseless transition rates. 
This is confirmed by a direct calculation.

To begin with I recall for reference the standard case $f(z) = z$. 
The double pole at $\tau = i\epsilon$ gives
\ba \label{usual R}
  R_+^{{\rm std}, In} = 0 \, , \quad R_-^{{\rm std}, In} = \frac{g^2 E}{2\pi} \, ,
\ea
The first result expresses the stability of the Minkowski vacuum in any inertial frame.

In the general case (\ref{W_f}), I note $z_k = \tau_k^2$ the poles of $W_f$, 
with $\tau_k \geq 0$ by convension.
Two cases must now be distinguished.
First, if ${\rm Im}(\tau_k) = 0$, we have
\ba \label{R padma}
  R_+^{In} &=& 0
\, , \quad 
  R_{-}^{In} = 
   e^{-\epsilon E} \, \frac{g^2 E }{2\pi} \sum_k \frac{\sin (E\tau_k)}{E\tau_k}
\qquad
\ea
As in the standart case, 
$R_{+} = 0$ because all the poles $i\epsilon \pm \tau_k$ are in the upper half 
complex plane. 
The simplest example of this class of propagators is (\ref{Gpadma}), 
for which we have only two simple poles in $\tau$, namely $\pm L + i\epsilon$.
For a uniformaly accelerated observer, the expression of the modified wightman function is
\ba \label{WwithL}
  {\cal W}(\tau) &=& - \frac{a^2}{16\pi^2} \, \frac{1}{\sinh^2(a\tau/2) - (La/2)^2}
\ea
and the transition rates become
\ba 
  R_{\pm}^{ua} &=&  R_{\pm}^{{\rm std},ua} \times \left( \sum_k 
   \frac{{\rm sinc}(E\rho_k)}{{\rm shc}(a\rho_k)} \right) \, ,
\ea
where $\rho_k = \frac{2}{a} {\rm argsh}(\frac{a\tau_k}{2})$.
I note $R_{\pm}^{\rm std}$ the rates given by field theories with standard propagators
\ba \label{Rstd}
  R_{\pm}^{{\rm std},ua} = \pm \frac{g^2E}{2\pi} \frac{1}{e^{\pm 2\pi E/a} - 1}
\ea

In the second case, where $Im(\tau_k)\neq 0$, we must have $Re(\tau_k)=0$ in order to 
obtain real-valued transition rates. 
The simplest propagator in this class is (\ref{Gparker}), with $\tau_k = \pm iL + i\epsilon$.
If we note $\tau_k = i \bar \tau_k +i\epsilon$, we obtain
\ba \label{stupid R}
  R_- &=& - \frac{g^2}{4\pi} e^{-\epsilon E} \sum_k \frac{e^{-E\bar \tau_k}}{\tau_k} 
\nonumber \\
  R_+ &=& - \frac{g^2}{4\pi} e^{+\epsilon E} \sum_k \frac{e^{-E\bar \tau_k}}{\tau_k}   
\ea
These expressions are clearly meaningless, if only because 
the transition rates diverge as $\tau_k \to 0$.
The expression for uniformly accelerated detectors 
are of course just as senseless as (\ref{stupid R}).
The authors of \cite{Gparker} argued that "to produce a physically 
sound result one must necessarily substract the naive inertial condibution", 
for the transition rates of accelerated as well as inertial detectors.
This is of course wrong since it amounts to remove the double pole $i\epsilon$
of the Wightman function, which encodes both the stability of the vacuum, 
see (\ref{usual R}), and
the causal properties of the theory via the identity (\ref{G_c}).
Considering the standard rate $R_-$ illustrates how erroneous is their logic
(they apparently only considered the rate
$R_+$, which vanishes for inertial detectors in the standard case):
in the inertial case $R_-^{{\rm std},In} \neq 0$ 
while their prescription gives a vanishing rate, and 
in the case of uniformly accelerated detectors, the rates are equal:
\begin{widetext}
\ba
 R_{-\, {\rm ref} [14]}^{\rm ua} \equiv R_-^{{\rm std},ua} - R_-^{{\rm std},In} 
 &=& g^2 \int_{-\infty}^{+\infty}\!\!d\tau \, e^{iE\tau} \,
\left\{ \frac{a^2}{4\sinh^2\frac{a}{2}(\tau-i\epsilon)} - \frac{1}{(\tau-i\epsilon)^2}  \right\}
\nonumber \\
  &=& g^2 \int_{-\infty}^{+\infty}\!\!d\tau \, e^{iE\tau} \,\left\{ 
 \sum_{n=-\infty}^{+\infty}\frac{1}{(\tau-i\epsilon - in\frac{2\pi}{a})^2} 
 - \frac{1}{(\tau-i\epsilon)^2}  \right\} 
\nonumber \\
  &=& \frac{g^2E}{2\pi} \,\sum_{n=1}^{\infty} e^{-n2\pi E/a}
\nonumber \\
  &=& e^{-\frac{2\pi E}{a}} \,  R_{-}^{{\rm std},ua} 
=  R_{+}^{{\rm std},ua}  \qquad ({\rm wrong})
\ea
\end{widetext}
As already explained, the error was introduced precisely 
by substracting the pole $\tau = i\epsilon$ of the 
thermal propagator.

\section{Euclidian propagators}
\label{sec:other}

Models are sometimes defined from the Euclidian propagator. 
An example found in the literature is \cite{Spallucci}
\ba \label{Euc1}
  G_E(p^2) = \frac{e^{-L^2 p_E^2}}{p_E^2 + m^2}
\ea
where $p_E^2 = {\bf p}^2 + p_4^2$ 
is the squared norm for the Euclidian scalar product.
The least one should require is that to (\ref{Euc1}) corresponds 
a well defined Feynman propagator in Minkowski space, since spacetime
is not Euclidian but Lorentzian, and Euclidian propagators are therefore
only meaningful if they are the analytic continuation of a well defined
propagator in Minkowski space (Wick rotation).
In standard QFT, the Euclidian and Feynman propagators are related
by $G_F(p^2 -i\epsilon) = i G_E(p_E^2)$. We see that the
Lorentzian version of (\ref{Euc1}) is not integrable in frequency. 
In particular its Fourier transform does not exist.
This means that neither the spectral function, nor 
the retarded and advanced propagators exist.

The authors of \cite{Spallucci} introduced the propagator 
(\ref{Euc1}) as a worthy tentative to define a 
unitary and Lorentz invariant NCFT.
Their proof of unitarity consists of the following steps:
they first adapt the cutting rules to the present propagator,
calculate with them the imaginary part of the 
self-energy at one loop, and compare the result
with an independent calculation of the decay propability. 
This calculation is done in the Euclidian.
Crutially, the final expression of the amplitude does 
not present an exponential factor $\exp(-L^2 p_E^2)$ of the 
external Euclidian momentum, 
as one would expect naively. This is what allowed them to do the 
analytic continuation of the amplitudes $p_E \mapsto -i \omega$
thus finding well behaved scattering amplitudes
verifying the optical theorem and usual bounds.

My criticism about this procedure is that these calculations are mere formal manipulations
done in analogy with the ones of standard QFTs. Indeed the propagator is 
only defined in the Euclidian, and these calculations only make sense in 
the Euclidian, so the analytic continuation of the Euclidian amplitude 
is not the amplitude of a relativistic QFT, since the latter does not exist. 
Moreover, the absence of an exponential factor $\exp(-L^2 p_E^2)$ in this 
amplitude might be due to the simple kinematics of the process, 
and it might very well be found in 
multiparticle scattering processes.

\section{Conclusion}

These notes put forth a warning concerning {\it ad hoc} models of 
a minimal length, namely that basic properties of the models, to wit  
causality and unitarity, should be checked before it is used in calculations.
Thus for instance, how should one interprete a scattering amplitude in a model 
without a Feynman propagator or a particle intepretation?  
This is not to kill all initiative regarding this approach, 
but elementary care should be exercised.
A safer way to construct models would be, for instance, to start from 
a spectral density and calculate its propagator by Fourier-Laplace transform.
This would at least guarantee unitarity, by construction.

\begin{appendix}

\section{Spectral function}

Writing the Euclidian propagators (\ref{K to Gpadma}) and (\ref{K to Gparker}) 
\ba
  G_E^\pm &=& \frac{1}{4\pi^2} \frac{1}{x_E^2 \pm L^2}
\nonumber \\
  &=& \frac{1}{16\pi^2} \int_0^{\infty}\!\!\frac{ds}{s^2} \, e^{-(x_E^2 \pm L^2)/4s} 
\ea
gives by identification with (\ref{spectral}) and (\ref{G_F})
\ba \label{defrho}
  e^{\mp L^2/4s} = \int_0^{\infty}\!\!dm^2 \, \rho_\pm(m^2) \, e^{- (m^2 + i\epsilon)s}
\ea
Hence $\rho_\pm(m^2)$ is the inverse Laplace-Fourier transform of $e^{\mp L^2/4s}$.

Let us first quote the following theorem:
Let $f(t)$ be a continuous function of the interval $\left[ 0,\, \infty  \right)$
which is of exponential order, that is, for some $b \in \mathbb{R}$.
\ba
  \sup_{t>0} \vert f(t) \vert e^{-bt} < \infty. 
\ea
In this case, its Laplace transform $F(s) = \int_{0}^{\infty}\!\!dt \, e^{-st} f(t)$ 
exists for all $s>b$ and is unique.
Then, $f \geq 0$ if and only if
\ba
  (-1)^n F^{(n)} \geq 0
\ea
for all $n\geq 0$ and all $s>b$.

In our case, $F(s) = e^{\alpha/s}$. Assuming that the spectral density is
of exponential order (a property which one checks {\it a posteriori}), 
application of the previous theorem shows that:
i) the inverse of $e^{-L^2/4\sigma}$ takes negative values;
ii) the inverse of $e^{-L^2/4\sigma}$ is non negative.
This will be verified by the direct calculation.
 
Let us now proceed with the calculation, begining with  (\ref{Gpadma}): 
\ba \label{LFtransf}
  \rho_+(m^2;L^2) = \frac{1}{i2\pi} \int_{\gamma-i\infty}^{\gamma+i\infty}\!\!d\sigma \, 
  e^{m^2\sigma - L^2/4\sigma}  \, .
\ea
The constant $\gamma$ is strictly positive in order to avoid the essential 
singularity at $\sigma = 0$.
After the two following change of variables $\sigma = iL\tau/2m$ and 
$t = \ln(\tau)$ (branch cut along the negative real axis), one obtains
\ba \label{rho}
  \rho_+(m^2;L^2) = \frac{L}{4\pi m}\, \int_{{\cal C}}\!\!dt \,  e^{t + i(m+i\epsilon) L \cosh t}
\ea 
where ${\cal C}$ is a contour in the strip $-\pi < {\rm Im} t < 0$  
that runs along ${\rm Im} t = -\pi$ from $\infty$ to the imaginary axis, then 
runs along the imaginary axis towards the origin of the plane, avoids the latter and 
finally goes back to infinity along the real axis. Explicitly,
\begin{widetext}
\ba
  \rho_+(m^2) &=& \frac{L}{4\pi m}\, \Bigl\{ 
 i\int_{0}^{\pi}\!\!dy \, e^{-iy+i(m+i\epsilon)L \cos y}
+ \, 2{\rm Re} \int_{0}^{\infty}\!\!dx \, e^{x + i(m+i\epsilon)L \cosh x} \Bigr\} 
\ea
The first integral, which corresponds to the vertical line of ${\cal C}$, 
is purely imaginary,
so multiplied by $i$ it gives a real contribution to $\rho_-(m^2)$.
Writing $e^x = \cosh x + \sinh x$, the second term in brakets
evaluates to 
\ba
 \int_{0}^{\infty}\!\!dx \, e^{x + i(m+i\epsilon)L \cosh x}  
 = K_1(imL) + i\frac{e^{imL}}{mL} \qquad
\ea 
Using $K_1(imL) = - \frac{\pi}{2} {\cal H}_{1}^{(2)}(mL)$, 
and ${\rm Re} {\cal H}_{1}^{(2)} = J_1 $, we get
\ba
  \rho_+ =  \frac{L}{4\pi m}\, \left\{ 
  i\int_{0}^{\pi}\!\!dy \, e^{-iy+i(m+i\epsilon)L \cos y} 
 - \pi \left[ J_1(mL) + \frac{2}{\pi} \frac{\sin mL}{mL}\right] \right\}
\ea
The remaining integral times $i$ is a real number, so we write
\ba 
  i\int_{0}^{\pi}\!\!dy \, e^{-iy+i(m+i\epsilon)L \cos y} = \int_0^{\pi}\!\!dy \, 
  \sin y \, e^{i(m+i\epsilon)L \cos y} = 2{\rm Re} \int_0^1\!\!du \, e^{imL u} 
  = 2 \frac{\sin mL}{mL}
\ea
The final resul is thus
\ba
  \rho_+ = - \frac{L}{4m} J_1(mL)
\ea
which as shown previously takes negative values. In particular 
the limiting expression for $mL \ll 1$ is
\ba
  \rho_+(m^2) =
 -\frac{L^2}{8}\left\{ 1 + O\left( (mL)^2 \right) \right\}
\ea
Moreover using the  
tabulated integrals
$\int_0^{\infty}\!\!dx\, J_\nu(bx) = b^{-1}$ for ${\rm Re}(\nu) > -1, \, b>0$, 
we also get
\ba
  \int_0^{\infty}\!\!dm^2 \, \rho_-(m^2) = -\frac{1}{2} \neq 1
\ea
In conclusion, $\rho_+$ cannot be interpreted as a spectral distribution.

Reproducing the same steps with (\ref{Gparker}) we get
\ba \label{def rho+}
  \rho_-(m^2) &=& \frac{1}{i2\pi} \int_{\gamma-i\infty}^{\gamma+i\infty}\!\!d\sigma \, 
  e^{m^2\sigma + L^2/4\sigma}   
\nonumber \\
&=& \frac{L}{4\pi m}\, \int_{\cal C}\!\!dt \,
  e^{t + i(m+i\epsilon)L \sinh t} 
\nonumber \\
  &=&\frac{L}{4\pi m}\, \Bigl\{ 
 i\int_{0}^{\pi}\!\!dy \, e^{-iy+(m+i\epsilon)L \sin y}
+ \, 2{\rm Re} \int_{0}^{\infty}\!\!dx \, e^{x + i(m+i\epsilon)L \sinh x} \Bigr\} 
\ea
The second term in the curly brakets is also equal to 
\ba \label{rho+ intermed1}
 \int_{0}^{\infty}\!\!dx \,\left(\cosh x + \sinh x  \right) \, e^{ i(m+i\epsilon)L \sinh x} 
 &=& \frac{i}{mL}  
  - \int_{-i\frac{\pi}{2}}^{\infty - i\frac{\pi}{2}}\!\!dy \, 
  \cosh y \, e^{- (m+i\epsilon)L \cosh y}
\ea
where the change of variable $x=y-i\pi/2$ has been made.
The latter integral can be calculated by application of 
Cauchy's theorem, closing the contour to make a rectangle with the 
opposite edge along the real axis,
\ba \label{rho+ intermed1}
  \int_{-i\frac{\pi}{2}}^{\infty - i\frac{\pi}{2}}\!\!dy \, 
  \cosh y \, e^{- (m+i\epsilon)L \cosh y}
  = \int_0^{\infty}\!\!dx \, \cosh x \, e^{- (m+i\epsilon)L \cosh x} 
 + i \int_0^{\pi/2}\!\!dx \cos x \, e^{-mL\cos x} \, .
\ea
The first integral on the right hand side
is real and recognized as $K_1(mL)$, and the second integral 
is also real. Combining (\ref{def rho+})-(\ref{rho+ intermed1}) one obtains
\ba
  \rho_-(m^2;L^2) =  \frac{L}{4\pi m}\, \left\{  K_1(mL) + 
  i\int_{0}^{\pi}\!\!dy \, e^{-iy+(m+i\epsilon)L \sin y}
 \right\}
\ea
Since $i$ times the integral in the brakets is real, we can alternatively write 
\ba
  \rho_-(m^2;L^2) =  \frac{L}{4\pi m}\, \left\{  K_1(mL) + 
  \int_{0}^{\pi}\!\!dy \, \sin y \, e^{(m+i\epsilon)L \sin y}
 \right\} \geq 0 \,.
\ea
Although positive, this function is not integrable over the positive values of 
its argument $m$,
\ba \label{sum rho+}
  \int_0^{\infty}\!\!dm^2 \, \rho_-(m^2) = \infty
\ea
since $K_1(z) \sim z^{-1}$ for $z \to 0$ and $e^{z\sin y}$ is not integrable 
at $z\to \infty$ for the whole the range of integration $y\in \left[ 0,\, \pi \right]$.
Finally in the limit $mL \ll 1$ one has
\ba
  \rho_-(m^2) = \frac{1}{4\pi m^2} \left\{ 1 + 2mL 
  + \frac{(mL)^2}{2} \ln \left( \frac{mL}{2} \right) + L^2\left( 1 + O(mL) \right)  \right\}
\ea
and we see also here the term $1/mL$ responsible for the divergence of (\ref{sum rho+})
at the lower bound. Because of this, $\rho_-$ cannot be interpreted as a spectral function.
\end{widetext}

\end{appendix}


\begin{thebibliography}{9}

\bibitem{Garay} L. J. Garay,
Int. J. Mod. Phys. A\textbf{10}, 145 (1995).

\bibitem{Seiberg} A. Connes, M.R. Douglas, A. Schwarz,
JHEP \textbf{9802}, 003 (1998);
M.R. Douglas and C.M. Hull,
JHEP \textbf{9802}, 008 (1998);
N. Seiberg and E. Witten,
JHEP \textbf{9909}, 032 (1999).

\bibitem{NCFT-Efield} N. Seiberg, L. Susskind, and N. Toumbas,
JHEP \textbf{0006}, 021 (2000).

\bibitem{NCFTnonunitary} J. Gomis and T. Mehen,
Nucl.Phys.B\textbf{591}, 265 (2000);
L. Alvarez-Gaum\'{e}, J.L.F. Barb\'{o}n, and R. Zwicky,
JHEP \textbf{0105}, 057 (2001).

\bibitem{UVcompletion} A. Adams, N. Arkani-Hamed, S. Dubovsky, A. Nicolis, R. Rattazzi,
JHEP \textbf{0610}, 014 (2006). 

\bibitem{SpeedofSound} G.F.R. Ellis, R. Maartens, M.A.H. MacCallum, 
Gen. Rel. Grav. \textbf{39}, 1651 (2007).


\bibitem{bookWeinberg} S. Weinberg, 
{\it The Quantum theory of fields. Vol. 1: Foundations},
Cambridge, UK: Univ. Pr. (1995).

\bibitem{BrosEpstein} J. Bros and H. Epstein, 
Phys. Rev. D \textbf{65}, 085023 (2002).

\bibitem{DynPreffered} T. Jacobson and D. Mattingly,
Phys. Rev. D \textbf{64}, 024028 (2001).

\bibitem{Mattingly} D. Mattingly, 
Rev. Rel. \textbf{8}, 5 (2005).



\bibitem{quasilocal} A. Jaffe, 
Phys. Rev. \textbf{158}, 1454 (1967).
V.Ya. Fainberg and M.A. Soloviev, 
Ann. Phys. \textbf{113}, 421 (1978).
 

\bibitem{Padmanabhan} T. Padmanabhan,
Phys. Rev. Lett. \textbf{78}, 1854 (1997).

\bibitem{PadmaUnruh} K. Srinivasan, L. Sriramkumar, and T. Padmanabhan,
Phys. Rev. D \textbf{58}, 044009 (1998).

\bibitem{Gparker} I. Agullo, J. Navarro-Salas, G. J. Olmo, and L. Parker,
Phys. Rev. D \textbf{77}, 124032 (2008).


\bibitem{UnruhLV} D. Campo and N. Obadia,
arXiv:1003.0112 [gr-qc].

\bibitem{Spallucci} A. Smailagic, E. Spallucci, 
J. Phys. A \textbf{37}, 1 (2004), Erratum-ibid. A \textbf{37}, 7169 (2004).

\end{thebibliography}
\end{document}